\documentclass[%
 aip,
 amsmath,amssymb,
reprint,%
]{revtex4-1}

\usepackage{graphicx}% Include figure files
\usepackage{dcolumn}% Align table columns on decimal point
\usepackage{bm}% bold math
\usepackage[utf8]{inputenc}
\usepackage[T1]{fontenc}
\usepackage{todonotes}
\def\iu{\ensuremath{\mathrm{i}}}
\let\phi\varphi

\begin{document}

\preprint{AIP/123-QED}

\title[Bumps and Oscillons in Spiking Networks]{Bumps and Oscillons in Networks of Spiking Neurons}
% Force line breaks with \\

\author{Helmut Schmidt}
\email[]{hschmidt@cbs.mpg.de}
\thanks{corresponding author}
\affiliation{
Max Planck Institute for Human Cognitive and Brain Sciences, 
Stephanstrasse 1a, 
04103 Leipzig, 
Germany}

\author{Daniele Avitabile}

\affiliation{%
  Department of Mathematics,
  Vrije Universiteit (VU University Amsterdam),
  Faculteit der Exacte Wetenschappen,
  De Boelelaan 1081a,
  1081 HV Amsterdam, \\
  Mathneuro Team, 
  Inria Sophia Antipolis,
  2004 Rue des Lucioles,
  06902 Sophia Antipolis, 
  Cedex.
}

\date{\today}% It is always \today, today,

\begin{abstract}
  We study localized patterns in an exact mean-field description of a
  spatially-extended network of quadratic integrate-and-fire (QIF) neurons.
We investigate
conditions for the existence and stability of localized solutions, so-called
\emph{bumps}, and give an analytic estimate for the parameter range where these
solutions exist in parameter space, when one or more microscopic network parameters
are varied.
We develop Galerkin methods for the model equations, which enable numerical
bifurcation analysis of stationary and time-periodic spatially-extended solutions.
We study the emergence of patterns composed of multiple bumps, which are arranged in
a snake-and-ladder bifurcation structure if a homogeneous or heterogeneous synaptic
kernel is suitably chosen. Furthermore, we examine time-periodic, spatially-localized
solutions (\emph{oscillons}) in the presence of external forcing, and in autonomous, 
recurrently coupled excitatory and inhibitory networks. In both cases
we observe period doubling cascades leading to chaotic oscillations.
\end{abstract}

\maketitle

%\begin{quotation}
%The ``lead paragraph'' is encapsulated with the \LaTeX\ 
%\verb+quotation+ environment and is formatted as a single paragraph before the first section heading. 
%(The \verb+quotation+ environment reverts to its usual meaning after the first sectioning command.) 
%Note that numbered references are allowed in the lead paragraph.
%%
%The lead paragraph will only be found in an article being prepared for the journal \textit{Chaos}.
%\end{quotation}

%\section{Introduction}

\begin{quotation}

Spatially extended networks of spiking model neurons are capable of producing
spatio-temporal patterns that are observed experimentally in neuronal tissues.
An important tool in investigating such patterns are low-dimensional neural field
models, which describe the macroscopic dynamics of such networks.
Many neural field models are derived heuristically, and do not fully
describe the dynamics of the underlying network of spiking neurons.
We utilize a recently derived mean-field description for networks of
quadratic integrate-and-fire neurons, which yields an accurate description
of the mean firing rate and the mean membrane potential of the network.
Contrary to other neural field models, this model only contains nonlinearities
of the mean field variables up to quadratic order, which are amenable to
the development of Galerkin methods for the numerical approximation of the problem.
This allows us to study the bifurcation structure of stationary localized solutions, 
and of time-varying localized solutions up to the emergence of chaos.

\end{quotation}

\section{Introduction}

Localized states in neuronal networks, so-called bumps, are
related to working memory~\cite{compte00,wimmer14} and feature
selectivity~\cite{kim17}, whereby neurons encoding similar stimuli or features show
an increased firing rate for the duration of the related cognitive task. 
Neural fields are well-known coarse-grained models of spatio-temporal neuronal
activity~\cite{wilson73,nunez74,amari77}, capable of reproducing dynamic phenomena
found experimentally, such as traveling waves, temporal oscillations, and spatially
localized states~\cite{coombes05,bressloff12}. 
A challenge faced in the derivation of neural field models is to establish an
accurate
mean-field description of the spiking dynamics of the underlying microscopic neural network. Classical neural field models
recover the microscopic dynamics only in the limit of slow
synapses~\cite{ermentrout94}, and the derivation of neural mass or neural field models from networks of
spiking neurons is still an active area of
research~\cite{ostojic11,schaffer13,buice13,visser14,mattia16,schwalger17,augustin17,park18,qiu18}.
In addition, in neural fields the network firing rate is not an emergent quantity,
but rather the result of a modelling choice.

Some limitations can be overcome if the microscopic model is
a heterogeneous network of synaptically coupled $\theta$ or QIF neurons, subject to
random,
% identically, and independently 
Cauchy-distributed background currents.
%\cite{montbrio15,coombes16}. 
Recently, it has been shown that heterogeneous networks of $\theta$- and QIF
neurons admit an exact mean field description
\cite{luke13,montbrio15}, which has later been expanded to
spatially-extended networks~\cite{laing14,laing2015exact,esnaola17,byrne19}.
In the thermodynamic limit, the network admits an exact
mean field description in terms of the network mean firing rate and
voltage~\cite{montbrio15}, or in terms of a complex-valued order
parameter~\cite{laing14,coombes16}.

We study a network of $n$ quadratic integrate-and-fire neurons:
\begin{equation}
  \dot V_i = V_i^2 + \eta_i + J s_i, \qquad i =1,\dots,n,
  \label{eq:micro}
\end{equation}
where $V_i$ is the membrane potential of the $i$th neuron, $\eta_i$ an intrinsic
current, $s_i$ the synaptic input and $J>0$ 
%\da{$J>0$ done} 
a global coupling parameter.
The $i$th neuron emits a spike when $V_i$ reaches the firing threshold $V_{\theta}$,
and $V_i$ is reset immediately
to $V_r$. Following Reference~\cite{montbrio15}, we distribute $\{\eta_i \colon i =
1,\ldots,n\}$ according to
a Lorentzian distribution using the formula 
$\eta_i = \eta + \Delta \tan \left[ \pi/2 (2 j -n -1)/(n+1) \right]$, where $\eta$ is
the center and $\Delta$ is the half-width of the Lorentzian distribution,
respectively. An important difference in the model considered in the present paper is
that neurons are distributed in space, in a domain 
$\Omega = (-L/2,L/2]$, with
$L \gg 1$, at evenly spaced positions $\{
x_i = i\delta x - L/2 \colon i \in 1,\ldots, n\}$ % $x_i = i \Delta x -L/2$, with $\Delta x = L / N$, 
with $\delta x = L/n$.
We associate with each lattice point $x_i$ a random component
of the vector $\{\eta_j\}$, without repetitions.
The synaptic current received by a neuron is determined by the synaptic footprint as
follows
\begin{equation}
  s_i(t) = \frac{1}{n} \sum_{j=1}^n w(x_i,x_j) \sum_{k \colon t_j^k < t}
  \int_{-\infty}^t a(t-t') \delta (t' - t_j^k)\, \textrm{d}t',
  \label{eq:syninput}
\end{equation}
where $w(x,y)$ models the synaptic coupling strength between neurons from position
$y$ to position $x$ in the network,
and $t_j^k$ is the emission time of the $k^{th}$ spike of the $j^{th}$ neuron.
The kernel $a(t)$ represents synaptic activation in response to incoming spikes,
e.g. exponential synapses $a(t) = \mbox{e}^{-t/\tau_s} / \tau_s$ with synaptic time scale $\tau_s$~\cite{devalle17,ratas18}.
In the mean field description we will let $\tau_s \rightarrow 0$, i.e. $a(t) =
\delta(t)$. 
We note in passing that, as demonstrated for leaky integrate-and-fire
neurons~\cite{zillmer07}, 
the mean field derivation
and the limit $\tau_s \rightarrow 0$ do not commute.

A neural field model that describes without approximation the average
firing rate $r(x,t)$ and the average membrane potential $v(x,t)$
of the spatially-extended networks presented above has been developed
recently~\cite{esnaola17}:
\begin{equation}
  \begin{aligned}
  \partial_t r &= \frac{\Delta}{\pi} + 2rv, \\
  \partial_t v &= v^2 +\eta +Jw\otimes r - \pi^2 r^2,
  \end{aligned}
  \qquad x \in \Omega.
  \label{eq:model}
\end{equation}
This neural field model inherits the coupling parameter $J$ and the parameters
$\eta$
and $\Delta$ from the microscopic, spiking network.
The mean field description is exact in the limit $n \to \infty$ and $V_{\theta} =
-V_r \rightarrow \infty$. 
The spatial coupling, or synaptic footprint, is given by the integral operator
\begin{equation}
  [w\otimes r](x) = \int_{\Omega} w(x,y) r(y) \mbox{d} y, \qquad x \in \Omega.
\end{equation}
For the concrete calculations presented below, we will
assume $\Omega = \mathbb{R}$ or $\Omega = (-L/2,L/2] \cong \mathbb{S}$ with $L \gg 1$ (a
ring with large width). 
We use $\Omega = \mathbb{R}$ for the theoretical framework in Section~\ref{sec:stat} and the Hermite-Galerkin method, 
and $\Omega = (-L/2,L/2] \cong \mathbb{S}$ 
for the Fourier-Galerkin method
and the numerical integration of both the macroscopic and microscopic model
equations, as periodicity is enforced in this setting.
We will study the model with a variety of kernels but, unless
stated otherwise, we assume, with a small abuse
of notation, $w(x,y) = w(|x-y|)$ and
\begin{equation}\label{eq:expkernel}
  w(x) = \mbox{e}^{-|x|} - \frac{1}{4} \mbox{e}^{-|x|/2},
\end{equation}
when $\Omega = \mathbb{R}$, or take the $2L$-periodic extension of $w$ when $\Omega =
(-L/2,L/2] \cong \mathbb{S}$. Hence our default synaptic kernel will depend on the
distance between two points in
$\Omega$, and will have long-range inhibition and short-range excitation. With these
choices $w \otimes r$ is a convolution and $\int_\Omega w(y) \mbox{d}y = 1$.

\begin{figure}
 \includegraphics[width=0.5\textwidth]{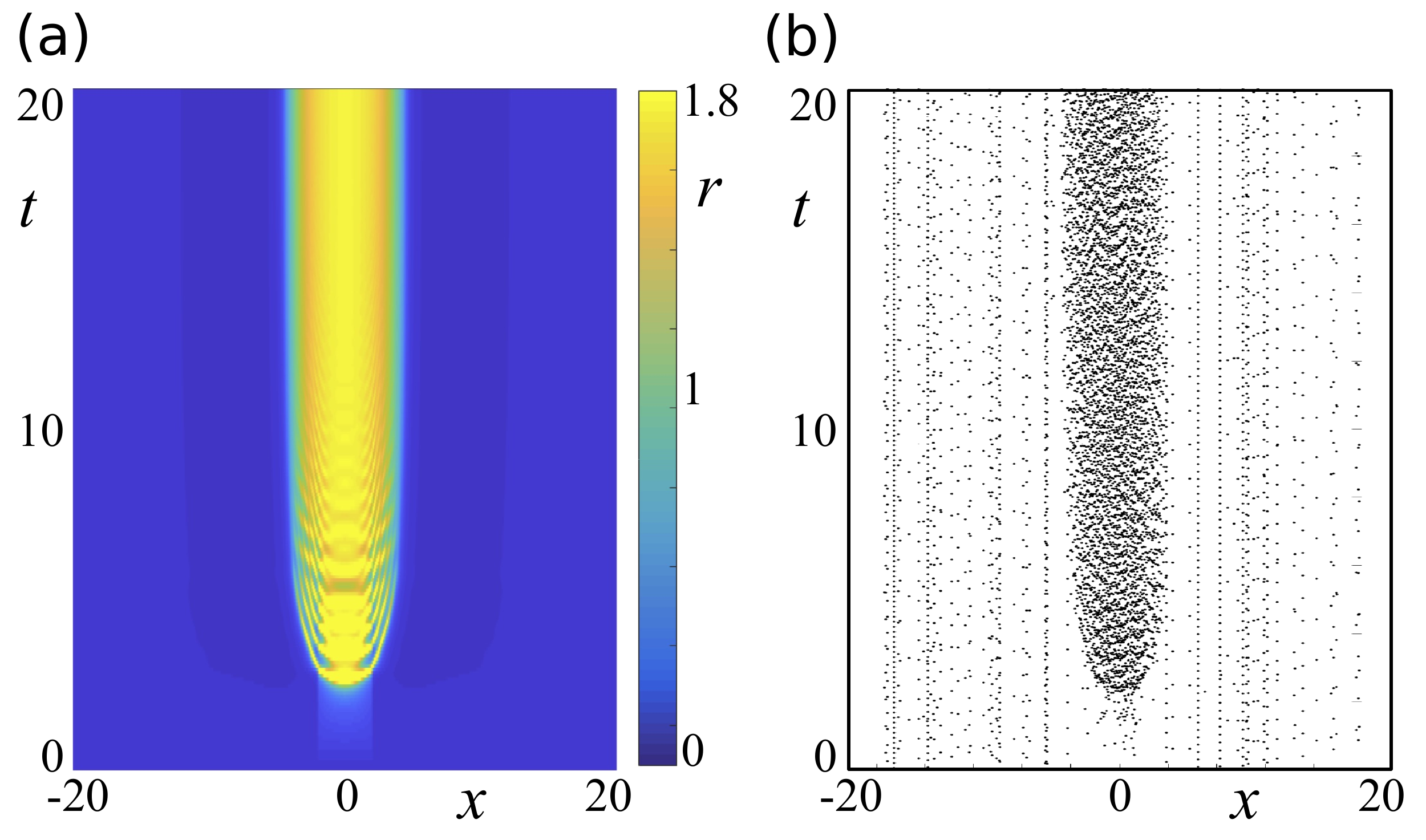}
 \caption{\label{fig1} 
   The formation of a stationary localized solution (bump) in (a) the QIF neural field
   model and (b) the corresponding network of spiking neurons. The bump solution is
   induced in the model with synaptic kernel \eqref{eq:expkernel}, by applying a
   localized transient current $I(x,t) \equiv 5$ if $(x,t) \in [-2.5,2.5] \times
   [0,5]$, and $I(x,t) \equiv 0$ otherwise. In
   (b) we show a rastergram of 300 of the $10^5$ neurons used in the simulation
   of the underlying spiking network. Parameters: $\Delta = 2$, $J = 15\sqrt{2}$,
 $\eta = -10$.}
 \end{figure}

This neural field model 
is related to mean field descriptions of networks of theta
neurons~\cite{luke13,laing14,coombes16,byrne19} and was obtained using the
Ott-Antonsen ansatz~\cite{ott08}. It retains the transient dynamics of
the microscopic network including spike synchrony, and has therefore a richer dynamic
repertoire than purely rate-based models~\cite{schmidt18}. 
The derivation of such neural field models is analogous to mean field approaches
for spatially extended networks of phase-coupled oscillators~\cite{kawamura14}.
An example of localized
solutions in this model is shown in Figure~\ref{fig1}a, alongside a
numerical simulation of the microscopic system of spiking neurons
in Figure~\ref{fig1}b.
%%%%%%%%%%%%%%%%%%%%%%%%%%%%%%%%%%%%%%%%%%%%%%%%%%%%%%%%%%%%%%%%%
%%%%%%%%%%%%%%%%%%%%%%%%%%%%%%%%%%%%%%%%%%%%%%%%%%%%%%%%%%%%%%%%%
 
The main aim of the present paper is to study spatiotemporal localized
patterns supported by this model, such as the one presented in Figure~\ref{fig1}.
Our investigation will be primarily numerical and therefore we will also introduce
several numerical schemes for the approximation of the QIF neural field model.
The paper is structured as follows: in Section~\ref{sec:stat} we discuss analytical
methods to study stationary solutions and their bifurcations; in
Section~\ref{sec:orth} we introduce the numerical methods used to perform numerical
bifurcation analysis of stationary and time-periodic localized structures, which are
presented in Sections~\ref{sec:loc} and \ref{sec:oscillons}, respectively; we make a
few concluding remarks in Section~\ref{sec:conclusion}.

\section{Stationary solutions}
\label{sec:stat}

Stationary states of Equation~\eqref{eq:model} are determined by the conditions $\partial_t
r=\partial_t v=0$. Bounded solutions
with $r(x) > 0$ satisfy
\begin{equation}
  0 = \frac{\Delta^2}{4\pi^2r^2} + \eta +Jw\otimes r - \pi^2 r^2,
  \quad
  v = -\frac{\Delta}{2\pi r}.
  \label{eq:stat}
\end{equation}
The model supports both uniform and non-uniform steady states, which we discuss
below in further detail.

\subsection{Spatially uniform states}
Solutions to \eqref{eq:stat} depend in general on $x$. Spatially-uniform solutions,
for which $J w \otimes r = J r$, satisfy the quartic equation
\begin{equation}
  r^4 -\frac{J}{\pi^2}r^3 -\frac{\eta}{\pi^2} r^2 - \frac{\Delta^2}{4\pi^4} = 0,
  \label{eq:statuni}
\end{equation}
which has the following four solutions,
\begin{equation} \label{eq:rSteadyStates}
  \begin{aligned}
  r_{1,2} = \frac{J}{4\pi^2} +\frac{1}{2}\sqrt{S} \pm \sqrt{-2p -S -2q/\sqrt{S}}, \\
  r_{3,4} = \frac{J}{4\pi^2} -\frac{1}{2}\sqrt{S} \pm \sqrt{-2p -S +2q/\sqrt{S}},
  \end{aligned}
\end{equation}
where $p$, $q$, and $S$ are given by
\begin{equation}
  p = -\frac{\eta}{\pi^2} -\frac{3}{8} \frac{J^2}{\pi^4}, \quad q = -\frac{J^3}{8\pi^6} - \frac{1}{2} \frac{J \eta}{\pi^4},
\end{equation}
\begin{equation}
  S = -\frac{2}{3}p +\frac{1}{3} \left( Q + R_0 / Q \right),
\end{equation}
with
\begin{equation}
  \begin{aligned}
  Q &= \left( \frac{1}{2}\left(R_1 + \sqrt{R_1^2 - 4 R_0^3} \right) \right)^{1/3}, \\
  R_0 &= \frac{\eta ^2}{\pi^4} - 3 \frac{\Delta^2}{\pi^4}, \\
  R_1 &= -2\frac{\eta ^3}{\pi^6} -\frac{27}{4} \frac{J^2 \Delta^2}{\pi^8} -
  18\frac{\eta \Delta^2}{\pi^6},
  \end{aligned}
\end{equation}
respectively. Physically-relevant solutions are positive and real, and an inspection of the
equations above reveals that $r_4$ must be discarded, and the system admits either
$1$ or $3$ homogeneous steady states.
At sufficiently small (large) $\eta$ only one stable fixed point exists, represented
by $r_3$ ($r_1$); also, there exists an interval in parameter space where
the stable solutions $r_1$, $r_3$ coexist with $r_2$, which is unstable. The
conclusions presented above justify the bifurcation diagram found in
References~\cite{montbrio15,esnaola17}, and reported in
Figure~\ref{fig:steadyStateHamiltonian}a.

Loci of saddle-node bifurcations in the $(\eta,J)$-plane can be found by setting
$\mbox{d} \eta / \mbox{d} r = 0$ in the first equation
in~\eqref{eq:stat} which, combined with~\eqref{eq:statuni} yields a parameterization
in $r$
\begin{equation}
  \begin{aligned}
    \eta_{sn} &= - \pi^2 r^2 - \frac{3 \Delta^2}{4 \pi^2 r^2}, \\
    J_{sn} &= 2\pi^2 r + \frac{\Delta^2}{2\pi^2 r^3},
  \end{aligned}
\end{equation}
or, more explicitly,
\begin{equation}
  \begin{aligned}
    J_{sn} & = \sqrt{2 \pi^2} \sqrt{ \!-\eta_{sn} \pm \!\sqrt{\eta_{sn}^2
    \!-\!3\Delta^2}} \\
    & + \frac{\sqrt{2\pi^2} \Delta^2}{\left(\!-\eta_{sn} \pm \!\sqrt{\eta_{sn}^2 \!-\!3 \Delta^2} \right)^{3/2} },
  \end{aligned}
  \label{eq:saddleetaJ}
\end{equation}
where $\pm$ denote two bifurcation branches of saddle-node bifurcation which collide
at a cusp
\begin{equation}
  \left(\eta_{c}, J_{c} \right) = \left( -\sqrt{3}\Delta, \frac{4\pi}{3} \sqrt{2\sqrt{3}\Delta} \right).
\end{equation}

\subsection{Turing bifurcations}
A first step towards the construction of heterogeneous steady states is the
determination of Turing bifurcations, which mark points in parameter space where
a spatially uniform solution becomes unstable to spatially periodic patterns. We
remark that it is known that spatially-extended networks of QIF or $\theta$
neurons display this instability~\cite{esnaola17,byrne19}, and here we present an
analytic determination of the loci of such bifurcation in parameter space.
Turing bifurcations of a homogeneous steady state $(r,v)$ can be identified
by linear stability analysis of the model equations in Fourier space,
which results in the following eigenvalue problem:
\begin{equation*}
 \lambda (k)
  \begin{pmatrix}
    \tilde r \\ \tilde v
  \end{pmatrix} =
  \begin{pmatrix}
    2v & 2r \\  J \hat w (k) - 2 \pi^2 r & 2v
  \end{pmatrix}
  \begin{pmatrix}
    \tilde r \\ \tilde v
  \end{pmatrix}
  := \hat A(k) 
  \begin{pmatrix}
    \tilde r \\ \tilde v
  \end{pmatrix},
  \label{eq:turingstab}
\end{equation*}
where $\hat w (k)$ is the Fourier transform of the connectivity kernel. A sufficient
condition for a Turing bifurcation is the existence of a critical wavenumber $k_c>0$
for which $\det A(k_c)=0$, which yields
\begin{equation}
  r_T \!=\! \frac{1}{\pi}\sqrt{ \!- \frac{ \eta_T \hat w (k_c)}{2 (2\!-\!\hat w
      (k_c)} \!\pm\! \frac{1}{2} \sqrt{ \frac{\eta_T^2 \hat w(k_c)^2}{(2\!-\!\hat w(k_c))^2} \!-\!\frac{(2\!+\!\hat w(k_c))\Delta^2}{2-\hat w(k_c)} } },
  \label{eq:turingr}
\end{equation}
and
\begin{equation}
  J_T = \frac{1}{\hat w (k_c)} \left( \frac{\Delta^2}{2 \pi^2 r_T^3} + 2 \pi^2 r_T \right).
  \label{eq:turingJ}
\end{equation}
Combining \eqref{eq:turingr} and \eqref{eq:turingJ} results in an equation for the
loci of the Turing bifurcation in the $(\eta,J)$-plane. As $k_c
\rightarrow 0$ the resulting equation recovers \eqref{eq:saddleetaJ}, since $\hat
w(k_c) \rightarrow 1$. This analytic result agrees well with the numerical
calculations of these loci, which will be presented further below. 

\subsection{Spatial dynamical system}
%%%%%%%%%%%%%%%%%%%%%%%%%%%%%%%%%%%%%%%%%%%%%%%%%%%%%%%%%%%%%%%%%
 \begin{figure*}
\includegraphics[width=0.9\textwidth]{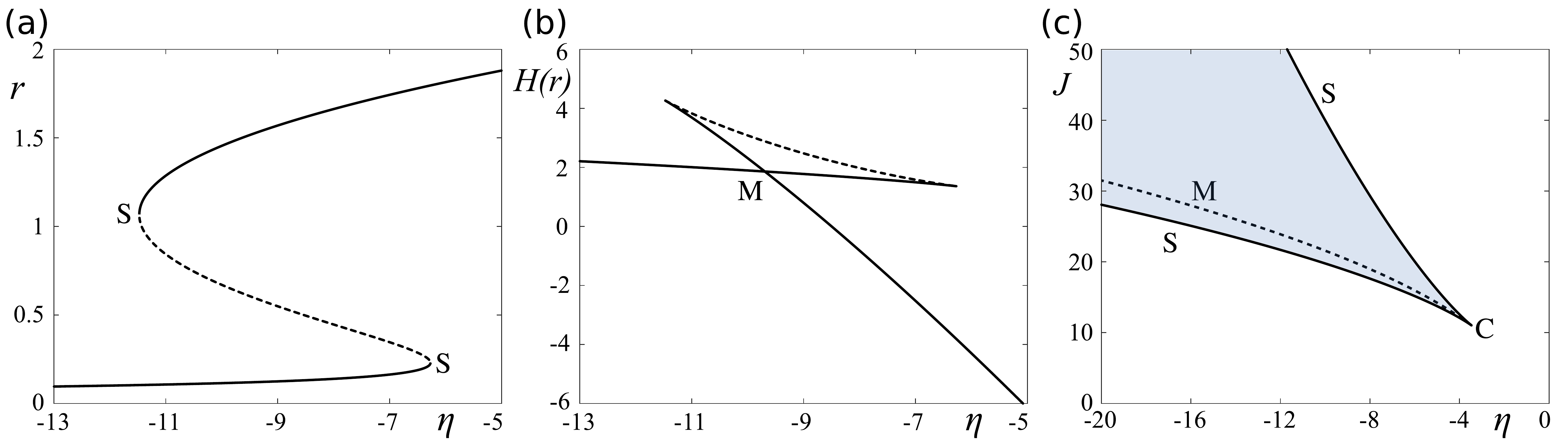}
 \caption{\label{fig2} (a) Solution branches of uniform solutions around the bistable
   regime with saddle-node bifurcations (S). (b) The associated conserved quantity $H$, as defined in \eqref{eq:HDef},
   showing a Maxwell point (M) at $\eta_M \approx -9.69$. (c) A plot of the bistable region of homogeneous states (shaded) 
   on the $(J,\eta)$-plane, delimited by saddle-node bifurcations (S), emanating 
   from a cusp (C). We also show the locus of Maxwell points (M) on the plane. Parameters: $J = 15\sqrt{2}$,
 $\Delta = 2$.}
 \label{fig:steadyStateHamiltonian}
 \end{figure*}
%%%%%%%%%%%%%%%%%%%%%%%%%%%%%%%%%%%%%%%%%%%%%%%%%%%%%%%%%%%%%%%%%
After studying uniform and spatially-periodic steady states, we
move to the construction of localized steady states supported by the models.
One strategy to study localized stationary states in nonlinear models posed on
$\mathbb{R}$ is to construct solutions to boundary-value problems derived from the
model's steady state equations
\cite{champneys98,burke07,burke07b,knobloch05,beck09}. With this
approach, localized steady states correspond to homoclinic orbits of a
dynamical system in which $x$ plays the role of time (hence the term
\textit{spatial dynamics}). 

In this section we make some preliminary considerations on the spatial dynamics of
steady state solutions to \eqref{eq:model}, although we do not explicitly study the
associated spatial-dynamical system, as we will construct our solutions numerically
in the following sections. Using the positions
\begin{align}
  & u(r) = -\frac{\Delta^2}{4\pi^2r^2} - \eta + \pi^2 r^2 = -v(r)^2 - \eta + \pi^2 r^2,
  \label{eq:defu} \\
  & f(u(r)) = J r,
   \label{eq:deffu}
\end{align}
the steady state equation~\eqref{eq:stat} is recast as
\begin{equation}
  0 = -u + w \otimes f(u).
  \label{eq:statamari}
\end{equation}
We note that Eq.~\eqref{eq:statamari} is formally equivalent to the Amari steady state
equation~\cite{amari77}; in Amari's theory $u$ represents the voltage, whereas in
this case $u$ combines the steady state's voltage 
and rate, and scales as $u \sim -v^2 - \eta$ for small $ r$ and $u \sim 
\pi^2 r^2 - \eta$ for large
$r$, respectively.

Importantly, the identification with the Amari equation allows us to use spatial
dynamics to characterize localized steady state
solutions~\cite{laing02,coombes03,elvin10,avitabile15}.
The Fourier transform of $w$ is of the form $\hat w(k) = \Pi_1(k^2)/\Pi_2(k^2)$, with
$\Pi_i$ being a polynomial of order $i$, if $w$ is the biexponential kernel~\eqref{eq:expkernel}.  
Hence, the integral kernel
can be regarded as Green's function of a differential operator.
In particular, the bi-exponential kernel~\eqref{eq:expkernel}
leads to the differential equation
\begin{equation}
  u'''' -\frac{5}{4} u'' + \frac{1}{4} u -  \frac{1}{4} f(u)+\frac{7}{4} [f(u)]'' = 0,
  \label{eq:amariode}
\end{equation}
where a prime denotes differentiation with respect to $x$. The equation above can be
cast as a 4D, first-order spatial dynamical system in the vector
$(u,u',u'',u''')$, which we omit here for brevity. To construct localized solutions
to \eqref{eq:model} we proceed in the same spirit as
\cite{champneys98,burke07,burke07b,knobloch05,beck09,laing02,elvin10,avitabile15}: to
each homogeneous steady state of
\eqref{eq:model} corresponds one value $r_j$ in \eqref{eq:rSteadyStates}, and hence
one value $u_j$ in \eqref{eq:defu}, and one constant solution $(u_j,0,0,0)$ to
\eqref{eq:amariode}; in addition, there exists a region in parameter
space where $r_1$ and $r_3$ coexist and are stable
(see also Figure~\ref{fig:steadyStateHamiltonian}a).
A localized steady
state of \eqref{eq:model} is identified with a bounded, sufficiently regular function $u \colon \mathbb{R}
\to \mathbb{R}$ which
satisfies \eqref{eq:amariode} with boundary conditions
\begin{align*}
  & \lim_{x \to -\infty} \big(u(x),u'(x),u''(x),u'''(x)\big) = (u_1,0,0,0),  \\
  & \lim_{x \to +\infty} \big(u(x),u'(x),u''(x),u'''(x)\big) = (u_3,0,0,0).
\end{align*}
Furthermore, we note that the quantity
\begin{align*}
  H(u,u',u'',u''',x) &= u''' u' -\frac{1}{2} (u'')^2 - \frac{5}{8} (u')^2  +
  \frac{1}{8} u^2 \\
  &- \frac{1}{4} \int^{u} f(z) \mbox{d} z \\
  &+ \frac{7}{4} \int^x [f(u)]''(z) u'(z) \mbox{d}z,
  \label{eq:hamilton}
\end{align*}
is conserved in the sense that, if \eqref{eq:amariode} holds, then
\[
  \frac{\mbox{d}}{\mbox{d}x}H(u(x),u'(x),u''(x),u'''(x),x) = 0.
\]
Therefore, we expect to construct a localized stationary state in a region of
parameter space where 
$H(u_1,0,0,0,0) = H(u_3,0,0,0,0)$. With a slight abuse of notation, we write this
condition in terms of the variable $r$, as $H(r_1) = H(r_3)$, where $H$ is
given by
\begin{equation}\label{eq:HDef}
  H(r) = -2\eta r -\frac{3}{2} J r^2 + \frac{4}{3} r^3.
\end{equation}
In analogy with the literature mentioned above, we called \emph{Maxwell points} the
values on the $(\eta,J)$-plane where the condition $H(r_1) = H(r_3)$ is met. We
display the Maxwell point for our standard parameter set in Figure~\ref{fig:steadyStateHamiltonian}b, 
and we plot the locus of Maxwell points and the
bistability region in Figure~\ref{fig2}c.

\section{Numerical Schemes}
\label{sec:orth}

As anticipated in the previous sections, stationary states beyond onset are computed
numerically, hence we present in this section several numerical schemes used in the
upcoming computations. In preparation for presenting the schemes, we rewrite the
model as an ODE on a function space. To simplify the notation we apply in this
section the scaling $r
\mapsto
r/\pi$, $J \mapsto \pi J$ to \eqref{eq:model}, and obtain 
\begin{equation}\label{eq:modelODE}
  \begin{aligned}
  \dot r &= \Delta + 2rv, \\
  \dot v &= \eta + v^2 - r^2 + Wr,
  \end{aligned}
\end{equation}
where $W$ is the integral operator defined as $(Wr)(x) = J (w \otimes r)(x) = J
\int_\Omega w(x,y) r(y) \mbox{d}y$.
In the system above we assume $r,v : \mathbb{R} \to L_\rho^2(\Omega,\mathbb{C})$
(also denoted by $L^2_\rho(\Omega)$), that is, at
each time $t$, $r(t)$ and $v(t)$ belong to a weighted Lebesgue space of
complex-valued functions defined on $\Omega$, with inner product 
\[
  \langle f, g \rangle_\rho = \int_\Omega f(x) g^*(x) \rho(x) \mbox{d} x,
\]
and norm $\Vert f \Vert_\rho = \langle f,f \rangle^{1/2}$. Note that the subscript
$\rho$ will be omitted when $\rho(x) \equiv 1$. We assume that, once complemented
with initial conditions, system \eqref{eq:modelODE} defines a well-posed Cauchy
problem on $L^2_\rho(\Omega) \times L^2_\rho(\Omega)$.

\subsection{Galerkin Schemes}
Galerkin schemes are derived by introducing a complete orthogonal basis
$\{\phi_i \colon i \in \mathbb{N}\}$ for the weighted space $L_\rho^2(\Omega)$, and
seeking an approximation in the $n$-dimensional subspace spanned by 
$\{ \phi_i \colon i \in \Lambda_n\}$, where $\Lambda_n$ is an index set with $n$
elements, as follows
\[
  r_n(x,t) = \sum_{i \in \Lambda_n} R_i(t) \phi_i(x), \quad 
  v_n(x,t) = \sum_{i \in \Lambda_n} V_i(t) \phi_i(x).
\]
A Galerkin scheme for \eqref{eq:modelODE} is then given by
\[
  \begin{aligned}
    & \langle \phi_i, - \dot r_n + \Delta + 2r_nv_n \rangle_\rho = 0, \\
    & \langle \phi_i, - \dot v_n + \eta + v_n^2 - r_n^2 + Wr_n \rangle_\rho = 0, 
  \end{aligned}
  \qquad i \in \Lambda_n,
\]
that is,
\[
  \begin{aligned}
    & \dot R_i = \alpha_i \Delta + 2 \sum_{j,k \in \Lambda_n} \gamma_{ijk} R_j V_k, \\
    & \dot V_i = \alpha_i \eta + \sum_{j \in \Lambda_n} \beta_{ij} R_j + 2 \sum_{j,k
    \in \Lambda_n} \gamma_{ijk} (V_j V_k - R_j R_k),
  \end{aligned}
\]
for $i \in \Lambda_n$, with coefficients given by
\[
  \alpha_i   = \langle \phi_i, 1 \rangle_\rho, \quad
  \beta_{ij} = \langle \phi_i, W\phi_j \rangle_\rho, \quad 
  \gamma_{ijk} = \langle \phi_i, \phi_j \phi_k \rangle_\rho.
\]

\subsubsection{Fourier-Galerkin Scheme}
When $\Omega = (-L/2,L/2] \cong \mathbb{S}$, the functions $r(t)$ and $v(t)$ are $L$-periodic. Therefore we choose the Fourier basis 
$\phi_j(x) = \exp( \iu j 2\pi x/L)$, $j \in \mathbb{Z}$, which is a complete
orthogonal basis for $L^2(\Omega)$. The index set for this case is $\Lambda_n =
\{-n/2,\ldots,n/2-1\}$ with $n$ even. Exploiting the trigonometric properties of the
Fourier basis, we obtain 
\[
  \alpha_i = 
  \begin{cases}
    L & \text{ if $i = 0$, } \\
    0 & \text{ otherwise, } 
  \end{cases}
  \quad
  \gamma_{ijk}
  =
  \begin{cases}
    L & \text{ if $i+j+k=0$,} \\
    0 & \text{ otherwise.} 
  \end{cases}
\]
In passing we note that $\beta_{ij}$ can also be expressed compactly, in terms of the
Fourier coefficients of the kernel $w$, if the operator $W$ is convolutional. In
addition, requiring $r$ and $v$ to be real-valued implies
$(R_i,V_i)=(R^*_{-i},V^*_{-i})$. We call this method the \emph{Fourier-Galerkin
scheme}.

\subsubsection{Hermite-Galerkin Scheme}
When $\Omega = \mathbb{R}$, a natural basis for the Galerkin scheme is given by the
Hermite polynomials 
\[
\phi_j(x) = H_j(x) = (-1)^j \exp(x^2) \frac{\mbox{d}^j}{\mbox{d} x^j} \exp(-x^2),
\]
which are a complete orthogonal set for $L^2_\rho(\Omega,\mathbb{R})$ with weight
$\rho(x) = \exp(-x^2)$. For this scheme $\Lambda_n = \{0,\ldots,n-1\}$. To avoid
problems with the numerical evaluations of $\phi_j$ for large $|x|$, we derive
an alternative scheme, which uses inner products with weight $\rho(x) \equiv
1$, as the Fourier Galerkin scheme. We seek a solution to
\eqref{eq:modelODE} in the form
\[
  r = R_0 + \tilde r, 
  \quad v = V_0 + \tilde v, 
\]
with $R_0$, $V_0$ constant in $x$, and $\tilde r, \tilde v \in
L^2(\Omega,\mathbb{R})$. This leads to the system
\[
  \begin{aligned}
  \dot R_0 &= \Delta + 2R_0 V_0,        \\
  \dot V_0 &= \eta + V_0^2 - R_0^2+ J R_0, \\
  \dot{\tilde r} &= 2R_0 \tilde v + 2V_0 \tilde r + 2 \tilde r \tilde v, \\
  \dot{\tilde v} &= 2V_0 \tilde v - 2R_0 \tilde r + \tilde v^2 - \tilde r^2 + W \tilde r, \\
  \end{aligned}
\]
in which the homogeneous background dynamics for $(R_0,V_0)$ is decoupled from
$(\tilde r,\tilde v)$, and follows the spatially-clamped QIF mean
field~\cite{montbrio15}.
Since the Hermite functions
\[
  \phi_j(x) = \exp(-x^2/2) H_{j-1}(x), \qquad j \in \mathbb{N}_{>0},
\]
are an orthogonal set for $L^2(\Omega,\mathbb{R})$, an approximation to $\tilde r,
\tilde v$ is sought in the space spanned by $\phi_j$, with $j \in \Lambda_n =
\{1,\ldots,n\}$, giving the scheme
  \begin{align*}
  \dot R_0  &= \Delta + 2R_0 V_0,        \\
  \dot V_0  &= \eta + V_0^2 - R_0^2+ J R_0, \\
  \dot{R}_i &= 2\sum_{j \in \Lambda_n} (R_0 V_j + V_0 R_j) 
             + 2 \sum_{j,k \in \Lambda_n} \gamma_{ijk} R_j V_k, \\
  \begin{split}
    \dot{V}_i &= \sum_{j \in \Lambda_n} [2V_0 V_j + (\beta_{ij} - 2R_0) R_j]
  \\ 
  &+ \sum_{j,k \in \Lambda_n} \gamma_{ijk}(V_j V_k - R_j R_k),
  \end{split} 
  \end{align*}
  for $i \in \Lambda_n$. We call this method the \emph{Hermite--Galerkin scheme}.

\subsection{Fourier Collocation Scheme}
A \emph{Fourier collocation scheme} can be derived in the case $\Omega = (-L/2,L/2]
\cong \mathbb{S}$. This method, which has been used in the past for Amari neural field
models \cite{rankin13,avitabile15} and the QIF neural field model \cite{esnaola17}, represents
$(r_n,v_n)$ by its values at the gridpoints $x_j=-L+2Lj/n$, $j \in \Lambda_n=\{
1,\ldots,n\}$, 
\[
  \begin{aligned}
    \dot R_i & = \Delta + 2R_i V_i, \\
    \dot V_i & = \eta + V_i^2 + R_i^2 + (Wr_n)_i, \\
  \end{aligned}
\]
and evaluates $(Wr_n)_i$ either with a quadrature rule or, more efficiently, with a
pseudospectral evaluation if $W$ is convolutional.

\subsection{Numerical considerations}
To the best of our knowledge, the methods presented above are novel, and we leave the
analysis of the
numerical properties of these schemes to a separate publication. The
calculations presented here have been tested against event-driven simulations of large
network of spiking neurons. We employ our schemes as follows: the Fourier collocation
scheme with $n=5000$ is generally used for time simulations, to obtain accurate
initial guesses for the continuation. However, we observed that time-periodic orbits
are reproduced with a similar accuracy by the Hermite--Galerkin scheme with just
$n=50$ modes, hence we select this scheme to continue periodic orbits. Finally,
we use the Fourier--Galerkin scheme with $n=200$ for bifurcation analysis of steady
states on large domains, when solutions are non-localized.

We compare the results of the QIF neural field model with the dynamics of the
underlying network of spiking neurons.
We integrate equation~\eqref{eq:micro} using the Euler method, with time step $\Delta t = 10^{-4}$.
The domain is chosen to be $\Omega = (-L/2,L/2]$ with periodic boundary conditions.
We choose $L=50$ and $5\times10^5$ model neurons, which ensures a good correspondence
to the neural field model ($\delta x = 10^{-4}$);
with $L=50$ the relative error of the normalization of the synaptic kernel, 
$\int_\Omega w(y) \mbox{d}y = 1$, is less than $10^{-5}$.
We note here that the microscopic description only matches the mean field description
if $\delta x$ is much smaller than the characteristic length scale of the synaptic kernel $w$.
The synaptic input to each neuron is computed with equation~\eqref{eq:syninput}.
We follow reference~\cite{montbrio15} 
in computing the synaptic integration
across a time window with $a_{\tau}(t) = \Theta(\tau-t)/\tau$, $\tau = 10^{-3}$, and 
in setting
$V_{\theta} = -V_r = 100$ with a refractory period of $2/V_i$ once the $i^{th}$
neuron has exceeded $V_{\theta}$.
The latter approximates the limit $V_{\theta} = -V_r \rightarrow \infty$.
The refractory period is rounded to the nearest multiple of $\Delta t$,
and after the refractory period $V_i$ is set to $-V_i$.
Rasterplots are generated using a subset of $10^3$ randomly chosen model neurons.

\section{Stationary localized solutions}
\label{sec:loc}
We use the numerical schemes presented in the previous section to study the
bifurcation structure of stationary localized solution to the QIF mean field model. We
initially study the model with our default excitatory-inhibitory
kernel~\eqref{eq:expkernel}, and then
show that a snaking bifurcation scenario is supported when the kernel is
switched to a homogeneous oscillatory kernel, or to a kernel with harmonic
heterogeneities, similarly to what is found for Amari neural field models.

\subsection{Local excitation, lateral inhibition kernel}
We set $w$ as in \eqref{eq:expkernel}, generate a stationary localized solution by
numerically integrating the model equations in time, and then implement the
Fourier--Galerkin scheme to continue the localized solutions in $\eta$, using AUTO.

\begin{figure}
\includegraphics[width=0.5\textwidth]{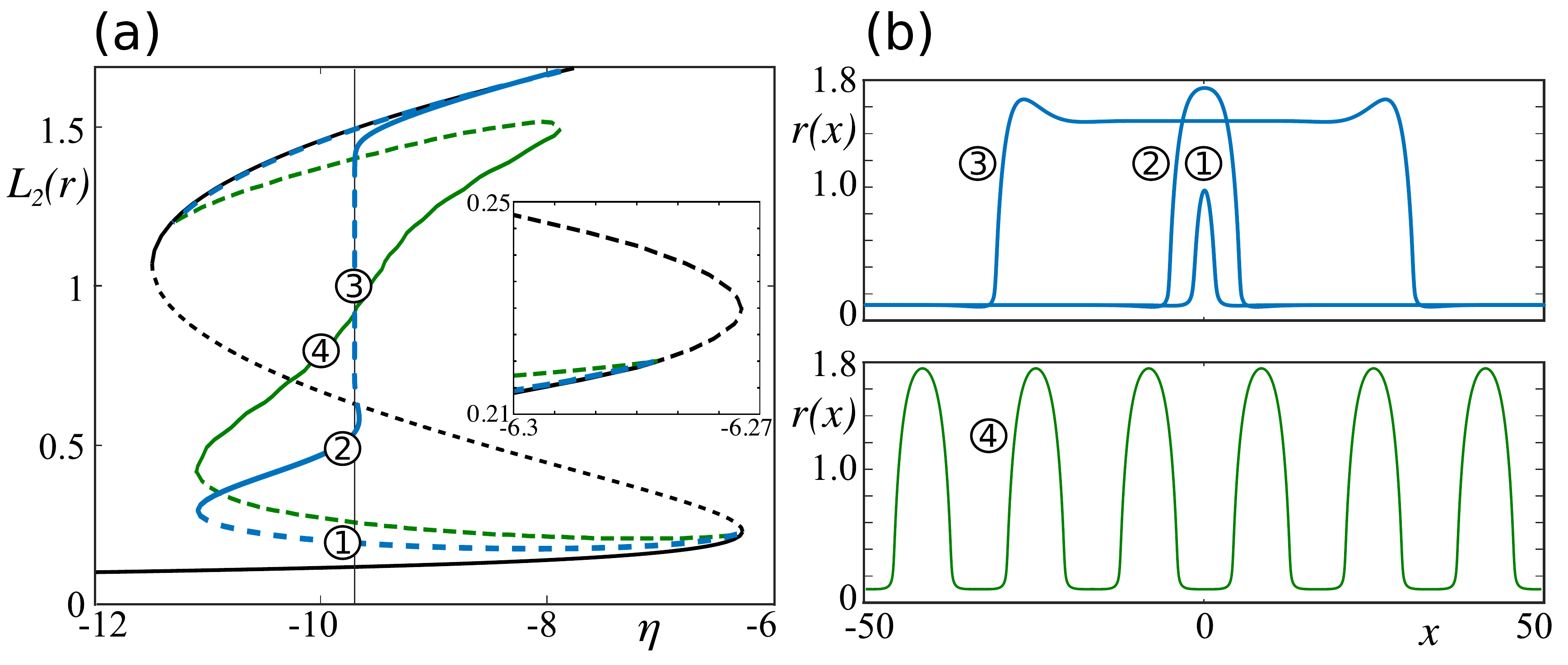}
 \caption{\label{fig6} (a) Bifurcation diagram in $\eta$ of localized solutions
   (blue), periodic solutions (green), and uniform solutions (black). (b) Exemplary
   profiles of unstable narrow (1), stable narrow (2), unstable wide (3), and stable periodic (4)
   localized solutions, close to the Maxwell point (vertical line in~(a)). Parameters:
   $\Delta = 2$, $J = 15\sqrt{\Delta}$. 
 }
 \end{figure}

In Figure~\ref{fig6} we show the bifurcation diagram of localized solutions. Across a
range of parameters, these
occur as a pair of one wide, stable solution and one narrow, unstable solution. 
The solution branch connects to the branch of uniform solutions at points where
Turing bifurcations occur, which also give rise to a branch of periodic solutions.
Using
the Fourier basis, it can be shown that the stable solution branch approaches the
Maxwell point asymptotically, and solutions grow wider, which resemble two
(stationary) interacting wave fronts. Because of the periodic boundary conditions,
the solution branch grows larger again and forms another stable/unstable solution
pair of locally low activity (not shown). The latter could be regarded as stationary
versions of \emph{traveling anti-pulses} reported in refs.~\cite{laing06,meijer14}. 

\begin{figure}
\includegraphics[width=0.45\textwidth]{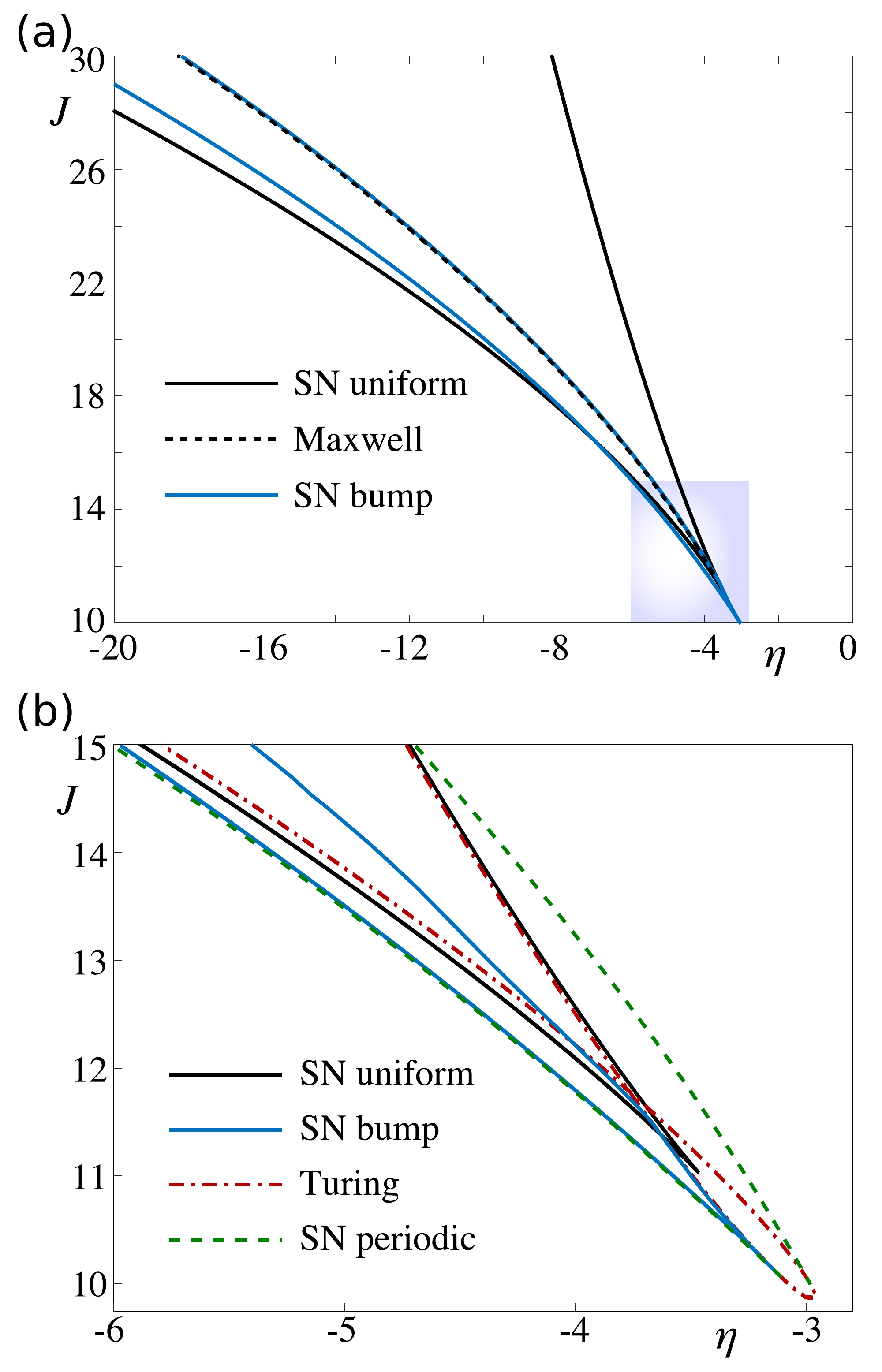}
 \caption{\label{fig7} (a) The parameter space in which stable localized solutions
 exist is delimited by loci of saddle-node bifurcations. They can be approximated by
 the
 saddle-node bifurcations of spatially uniform solutions, and the Maxwell point. (b)
 Inset of (a), showing additionally the loci of Turing bifurcations and
 saddle-node bifurcations of periodic solutions. The saddle-node bifurcations of the bump
 solutions form a cusp where the Turing bifurcation changes from supercritical to
 subcritical. 
 Parameters: $\Delta = 2$. }
 \end{figure}

Because stable solutions are of particular interest, we present a
two-parameter bifurcation diagram (Figure~\ref{fig7}a) of the saddle-node
bifurcations that delimit the branch of stable solutions.
As expected, the locus of saddle-node bifurcations 
of localized states enclose the Maxwell point. In
addition, the loci of saddle-node bifurcations of localized and uniform steady states meet at two separate
cusps, as shown in Figure~\ref{fig7}b. 

The bifurcation behavior of localized solutions described above is robust to changes
in coupling parameters but, as we shall see below, it is strongly affected by changes
in the kernel.

 %%%%%%%%%%%%%%%%%%%%%%%%%%%%%%%%%%%%%%%%%%%%%%%%%%%%%%%%%%%%%%%%%
 %%%%%%%%%%%%%%%%%%%%%%%%%%%%%%%%%%%%%%%%%%%%%%%%%%%%%%%%%%%%%%%%%

 %%%%%%%%%%%%%%%%%%%%%%%%%%%%%%%%%%%%%%%%%%%%%%%%%%%%%%%%%%%%%%%%%
 %%%%%%%%%%%%%%%%%%%%%%%%%%%%%%%%%%%%%%%%%%%%%%%%%%%%%%%%%%%%%%%%%

\subsection{Snaking with homogeneous kernel}
\label{sec:snakes}

\emph{Homoclinic snaking} is a phenomenon that describes the formation of multiple,
coexisting localized solutions in spatially-extended models. Steady states are
arranged in branches of intertwined snaking bifurcation diagrams, connected via
ladders~\cite{Champneys1998,knobloch05,burke07,lloyd08,avitabile10}.
Adopting the spatial-dynamics approach outlined above, localized solutions are
interpreted as homoclinic orbits to a fixed point. Snaking solution branches
correspond to symmetries of the problems,
which are broken along the ladder branches~\cite{burke07b,beck09}. This scenario is
not limited to PDEs, but have also been studied in the non-local Swift-Hohenberg
equation~\cite{morgan14}, as well as in neural field
models~\cite{laing02,coombes03,laing03,faye13,faye13b,rankin13,avitabile15}.  

\begin{figure}
\includegraphics[width=0.5\textwidth]{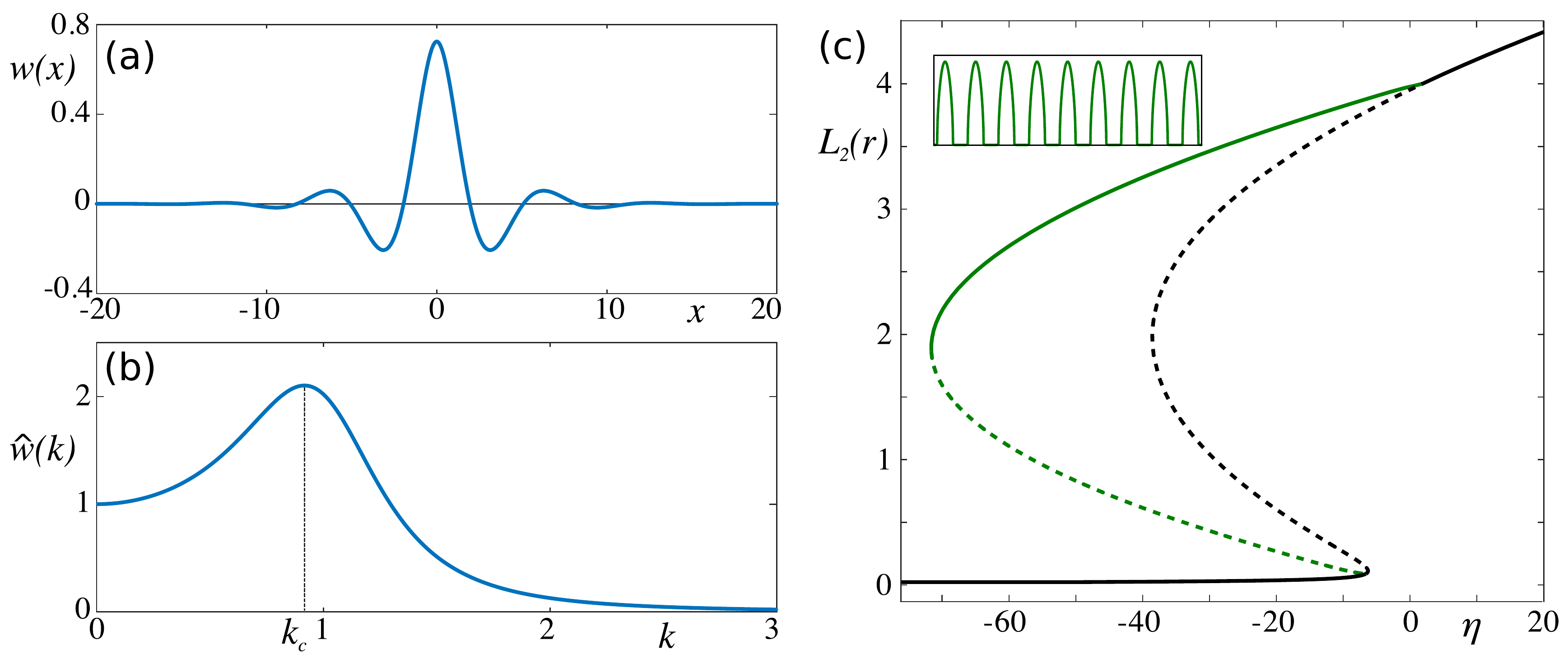}
 \caption{\label{fig8} (a) The damped-oscillatory kernel \eqref{eq:dampedKernel}. (b)
   Fourier transform of the kernel, which shows a maximum at a non-zero wave-number $k_c$.
   (c) As $\eta$ is varied, the branch of homogeneous steady states (black)
   undergoes a Turing bifurcation, from which a branch of periodic solutions with
   wavenumber $k_c$ emerges (green). Parameters: $J = 39$, $\Delta = 1$,
 $b=0.4$. }
 \end{figure}

In the simplest setting, localized snaking solutions are found in regions of
parameter space where there is bistability between a stationary homogeneous state and
a periodic state. In nonlocal neural fields, homoclinic snaking has been observed 
with
the following homogeneous damped-oscillatory kernel~\cite{elvin10}
\begin{equation}\label{eq:dampedKernel}
  w(x) = \frac{1+b^2}{4b}\,\mbox{e}^{-b|x|} \left( b \sin|x| + \cos x \right),
\end{equation}
which we now adopt also for the QIF neural field model. This kernel leads to a
sub-critical Turing bifurcation of the lower stable branch of
uniform solutions, from which an unstable branch of spatially-periodic solutions
emerges. This branch undergoes a saddle-node bifurcation, where spatially-periodic
solutions become stable. Eventually, the branch connects to the
upper stable branch of uniform solutions, see Figure~\ref{fig8}.

 %%%%%%%%%%%%%%%%%%%%%%%%%%%%%%%%%%%%%%%%%%%%%%%%%%%%%%%%%%%%%%%%%
 %%%%%%%%%%%%%%%%%%%%%%%%%%%%%%%%%%%%%%%%%%%%%%%%%%%%%%%%%%%%%%%%%
 
\begin{figure}
\includegraphics[width=0.5\textwidth]{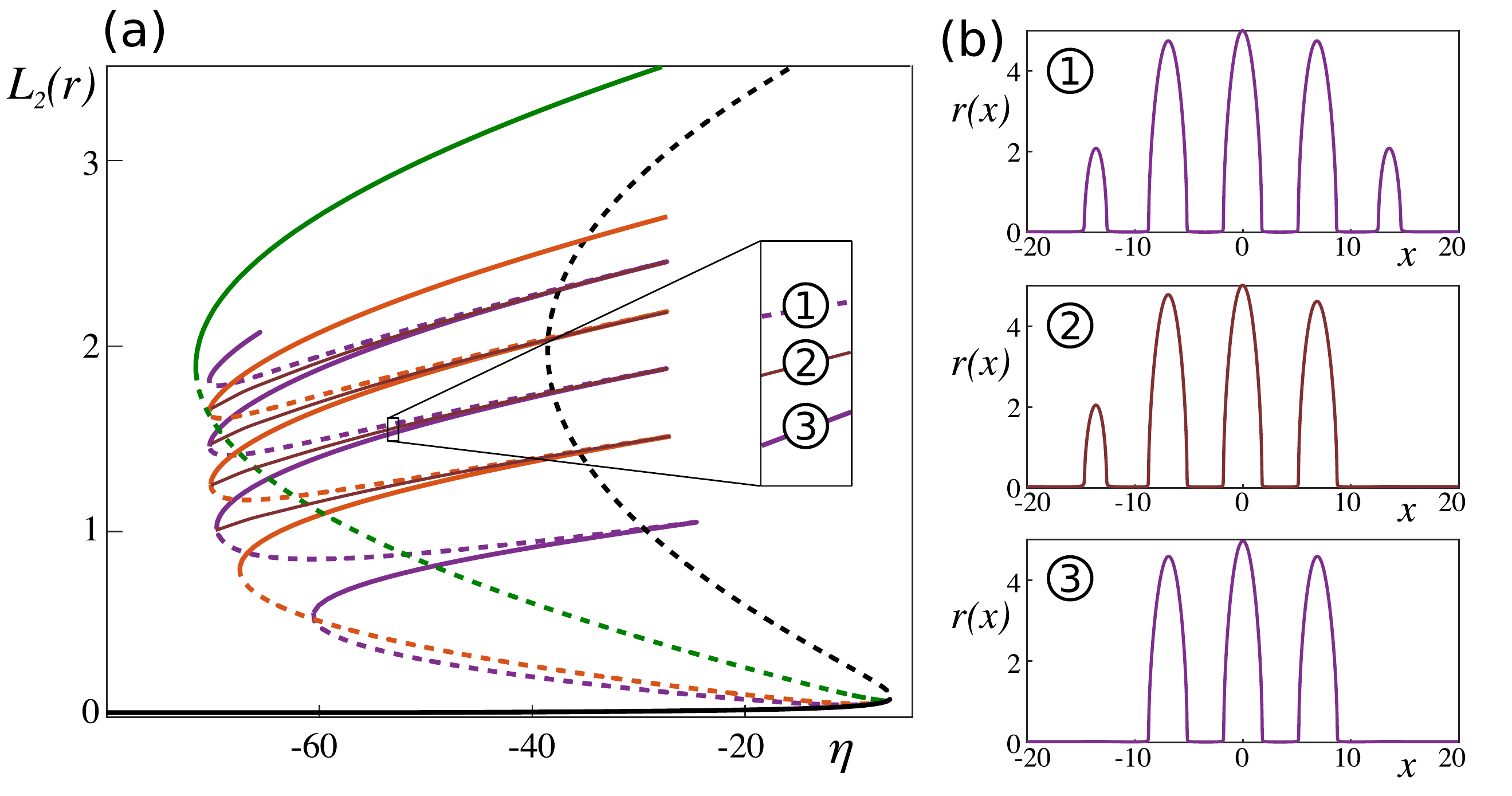}
 \caption{\label{fig9} (a) Snakes-and-ladders bifurcation scenario. (b) Representative
   solutions for branches of symmetric (1,3) and asymmetric
   solutions (2). Parameters: $J = 39$, $\Delta = 1$, $b=0.4$.}
 \end{figure}

 As anticipated, spatially localized snaking solutions are found in this region of
 parameter space, and they are arranged in a
 typical snakes-and-ladders bifurcation structure, which is displayed in
 Figure~\ref{fig9}.
 
  %%%%%%%%%%%%%%%%%%%%%%%%%%%%%%%%%%%%%%%%%%%%%%%%%%%%%%%%%%%%%%%%%
 %%%%%%%%%%%%%%%%%%%%%%%%%%%%%%%%%%%%%%%%%%%%%%%%%%%%%%%%%%%%%%%%%

%

\subsection{Snaking with heterogeneous kernel}
It is known that snaking bifurcation scenarios can be triggered by heterogeneities in
the underlying evolution equations. Examples discussed in the literature include the
Swift--Hohenberg~\cite{kao2014}, Amari~\cite{avitabile15}, and
Ginzburg-Landau~\cite{ponedel16} equations. 
In neural field models, heterogeneities are naturally introduced via harmonic
perturbations of a homogeneous (distance-dependent) kernel, which break the
translational invariance of the
problem~\cite{bressloff01,kilpatrick08,schmidt09,coombes11}. In
Reference~\cite{avitabile15} we have shown that the following kernel leads to snaking
in the Amari model
\begin{equation}\label{eq:cosKernel}
w(x,y) = \frac{1}{2} \mbox{e}^{-|x-y|} (1 + a \cos( k y)),
\end{equation}
and we therefore investigate the effect of this kernel on the QIF neural field model.

\begin{figure}
\includegraphics[width=0.5\textwidth]{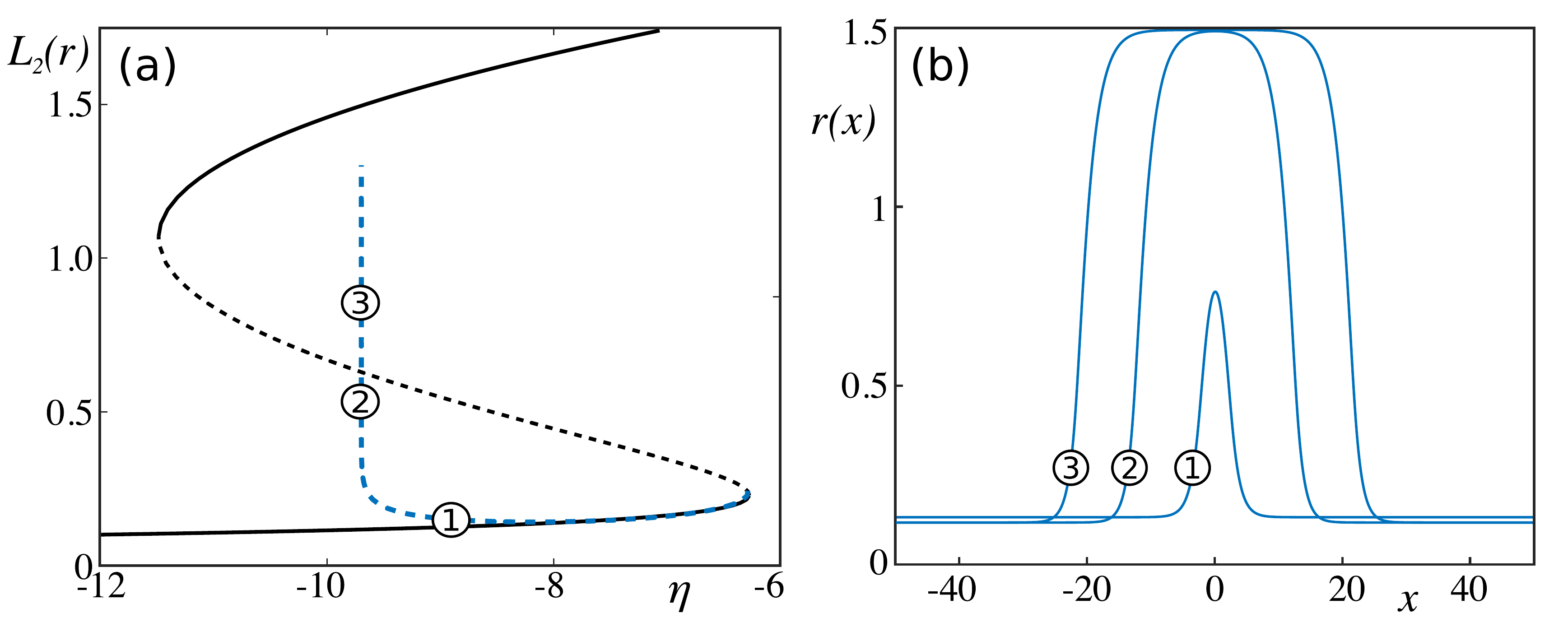}
 \caption{\label{fig10} (a) Bifurcation diagram of spatially uniform solutions, and
   of localized solutions generated with the exponential kernel \eqref{eq:cosKernel}
   with $a=0$. The lack of lateral inhibition results in the entire branch of
 solutions being unstable. (b) Representative solutions. Parameters: $J =
 15\sqrt{\Delta}$, $\Delta = 2$.}
 \end{figure}

\begin{figure}
\includegraphics[width=0.5\textwidth]{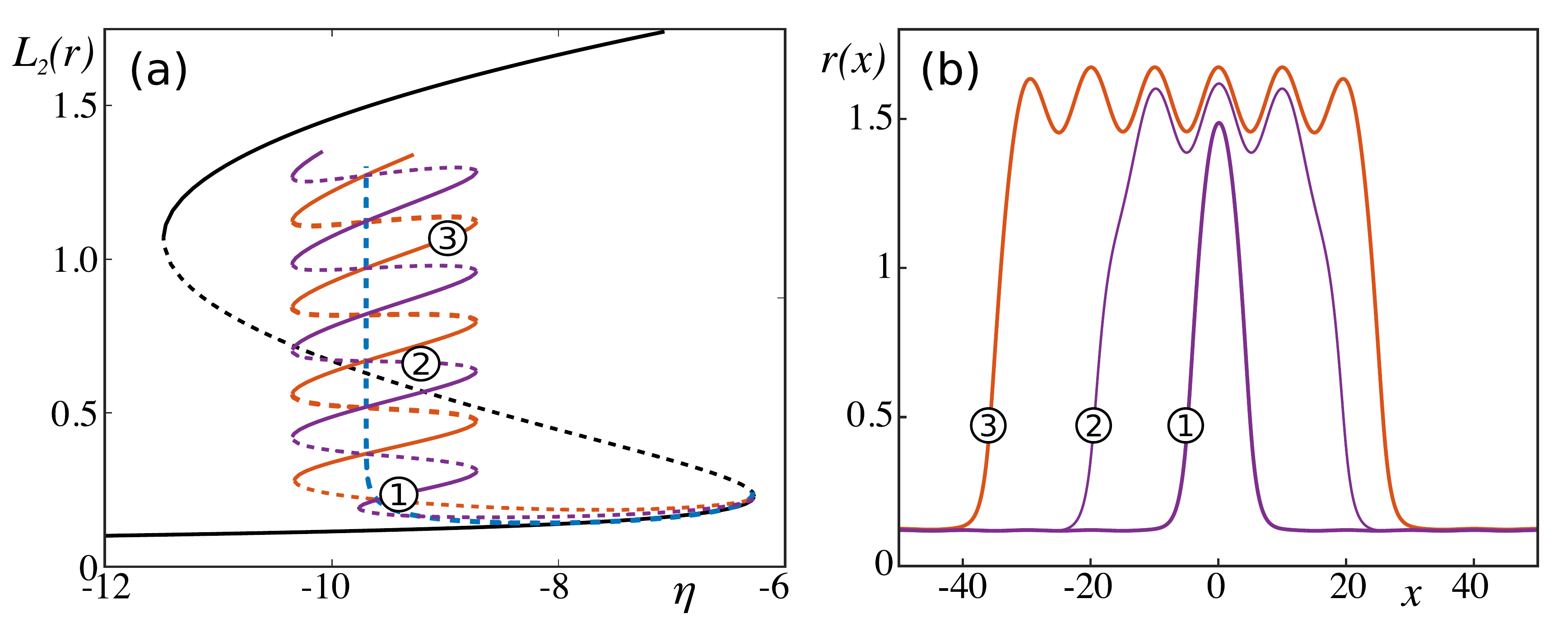}
 \caption{\label{fig11} Snaking induced by spatial periodic modulation
   of the exponential connectivity kernel \eqref{eq:cosKernel}. (a) Bifurcation
   diagram obtained for $a=0.1$ (orange and purple branches, ladders not shown) 
   compared with the one obtained for $a=0$ (blue). (b)
   Representative solutions. Parameters: $J = 15\sqrt{\Delta}$, $\Delta = 2$.}
 \end{figure}

In the absence of spatial forcing ($a=0$), a system with exponential connectivity
does not yield stable localized solutions (see Figure~\ref{fig10}).
%%%%%%%%%%%%%%%%%%%%%%%%%%%%%%%%%%%%%%%%%%%%%%%%%%%%%%%%%%%%%%%%%
%%%%%%%%%%%%%%%%%%%%%%%%%%%%%%%%%%%%%%%%%%%%%%%%%%%%%%%%%%%%%%%%%
In the presence of modulation, we find snaking branches that oscillate around the
branch obtained for $a=0$ (see Figure~\ref{fig11}). Furthermore, for small values of
$a$, the snaking width increases proportionally to the value of $a$ (not shown).
These findings indicate that the snaking phenomenon in the QIF neural field model is
entirely determined by the kernel choice, as in the Amari case.
%%%%%%%%%%%%%%%%%%%%%%%%%%%%%%%%%%%%%%%%%%%%%%%%%%%%%%%%%%%%%%%%%
%%%%%%%%%%%%%%%%%%%%%%%%%%%%%%%%%%%%%%%%%%%%%%%%%%%%%%%%%%%%%%%%%

%

 \section{Oscillons}\label{sec:oscillons}
Various nonlinear models including chemical, fluid-dynamical, and
particle systems, support time-periodic, spatially-localized states termed
\emph{oscillons} (see Reference~\cite{Knobloch2008g} and references therein). A
comprehensive theory for the existence and
bifurcation structure of such solutions is the subject of experimental, numerical,
and analytical investigations. We study oscillons in the QIF neural field model
in the two main settings where they are observed in other media: (i) a
non-autonomous setting, whereby oscillons emerge as the medium is subject to a
homogeneous, exogenous, time-periodic forcing; (ii) an autonomous setting, whereby
oscillons emerge spontaneously as one of the model parameters is varied.

\subsection{Oscillons induced by harmonic forcing}
We setup the QIF neural field model subject to a time-dependent, homogeneous, sinusoidal
forcing with frequency $\omega$,
\begin{equation*}
  \begin{aligned}
  \partial_t r &= \frac{\Delta}{\pi} + 2rv, \\
  \partial_t v &= v^2 + Jw\otimes r - \pi^2 r^2 + \eta + A \sin(\omega t),
  \end{aligned}
\end{equation*}
and cast it in the following, equivalent autonomous model formulation to perform
numerical bifurcation analysis
\begin{equation}
  \begin{aligned}
  \partial_t r &= \frac{\Delta}{\pi} + 2rv, \\
  \partial_t v &= v^2 + Jw\otimes r - \pi^2 r^2 + \eta + A \xi, \\
  \dot \xi   &= \xi   + \omega \zeta - (\xi^2 + \zeta^2) \xi, \\
  \dot \zeta &= \zeta - \omega \xi - (\xi^2 + \zeta^2) \zeta. \\
  \end{aligned}
  \label{eq:model_forced}
\end{equation}
Note that the numerical framework proposed here is applicable also if the forcing is
heterogeneous. 

In this setting we expect oscillons to emerge without bifurcation from a localized
steady state of the QIF neural field model with $A=0$, upon imposing a small-amplitude
forcing, $A \ll 1$. We therefore select the default kernel~\eqref{eq:expkernel},
set $\eta = -10$, for which the model with $A=0$ supports one stable (wide) and one
unstable (narrow) bump (see Figure~\ref{fig6}), and continue time-periodic solutions to
\eqref{eq:model_forced} in $A>0$ for $\omega = 4$, close to the network's resonant frequency~\cite{schmidt18}.
%
%%%%%%%%%%%%%%%%%%%%%%%%%%%%%%%%%%%%%%%%%%%%%%%%%%%%%%%%%%%%%%%%%
 \begin{figure*}
\includegraphics[width=0.9\textwidth]{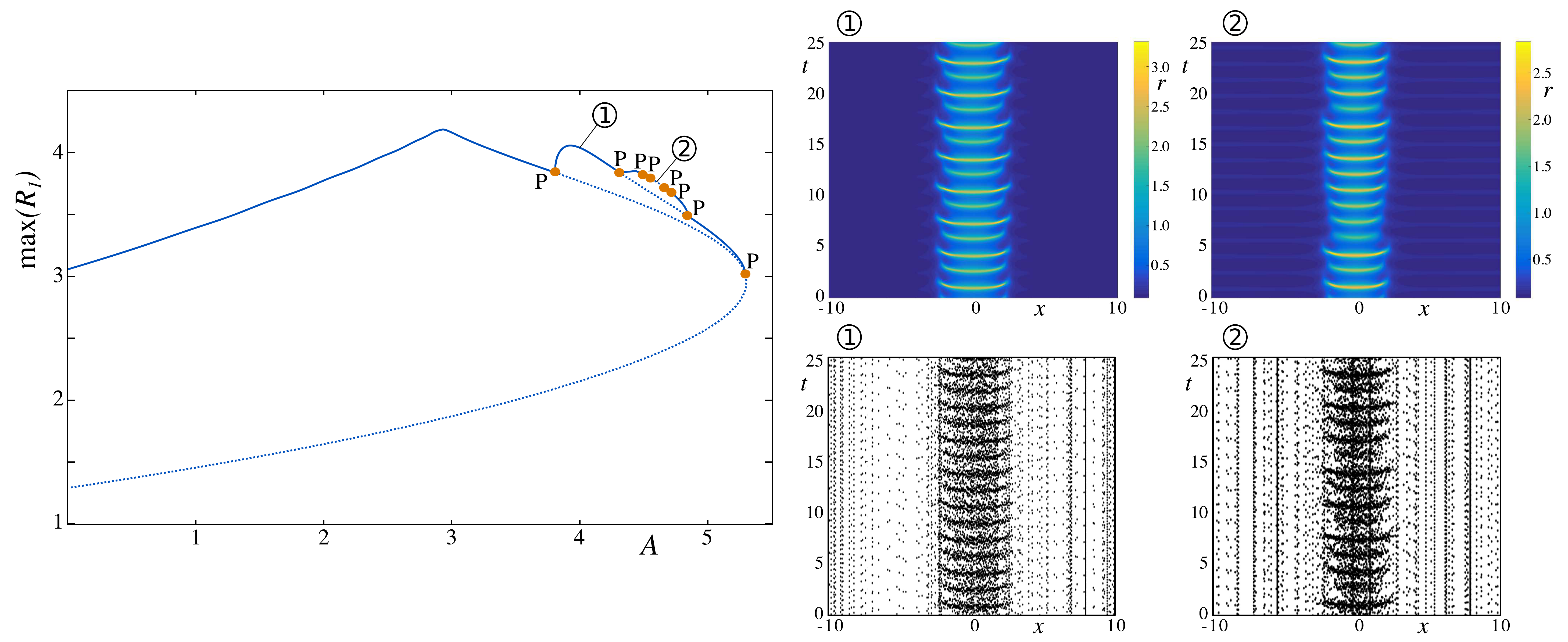}
 \caption{\label{fig12} Maximum values of $R_1(t)$ plotted against
 the amplitude of sinusoidal forcing. At $A\approx 3.8$ a period-doubling bifurcation
 occurs, which is the starting point of a period-doubling cascade leading to chaos at
 $A\lesssim 4.6$. 
 A period-doubled solution ($A=4$) and a chaotic solution ($A=4.6$) is shown.
 Parameters: $\Delta = 2$, $J = 15\sqrt{\Delta}$, $\eta =
 -10$, $\omega = 4$.}
 \end{figure*}
%%%%%%%%%%%%%%%%%%%%%%%%%%%%%%%%%%%%%%%%%%%%%%%%%%%%%%%%%%%%%%%%%

One stable and one unstable branch of oscillons emerge from $A=0$, as shown in
Figure~\ref{fig12}, and connect at a saddle-node bifurcation. The stable
branch undergoes a sequence of period-doubling bifurcations leading to chaos, and
examples of a period-doubled solution and a chaotic solution are shown in
Figure~\ref{fig12}, demonstrating the correspondence between the QIF neural field model and
the spiking network model.

In a recent study we have investigated the effect of periodic forcing
on a population of excitatory spiking neurons~\cite{schmidt18},
whose solutions correspond to the spatially uniform
states of the present model. In that context it was shown that a sufficiently
large forcing amplitude is able to suppress homogeneous oscillations. Here we report
that the same statement holds true for forced oscillons: no localized time-periodic
solution is found to the right of the saddle-node bifurcation in
Figure~\ref{fig12}, where the attractor is a spatially-homogeneous, time-periodic
state, which can be found by continuing in $A$ the low-activity
uniform steady state (not shown).

\subsection{Spontaneous oscillons in coupled networks of excitatory and inhibitory
neurons}

In the second scenario, oscillons occur in autonomous systems. 
Direct numerical simulations of reaction diffusion systems display oscillons
in the proximity of codimension-two Turing--Hopf bifurcations of the homogeneous
steady state~\cite{Vanag:2004hw,Vanag:2007ks}. Oscillons in these systems have
typically been observed as large-amplitude structures, hence they are conjectured to
form via a subcritical Hopf bifurcation of a heterogeneous, spatially-localized
steady state. This conjecture, however, has not yet been confirmed by numerical
bifurcation analysis which, in contrast to direct numerical simulations, allows to
track both stable and unstable states. 

Here we employ the Hermite--Galerkin scheme to
study the formation of oscillons in the QIF neural field model. As mentioned above, a necessary ingredient for
oscillons is the presence of oscillatory bifurcations. These bifurcations are
precluded in one-populations networks of QIF neurons but,
as we shall see,
are possible
in two-population models, therefore we turn our attention to the
following network of coupled excitatory and inhibitory populations
\begin{equation*}
  \begin{aligned}
  \dot r_e &= \frac{\Delta}{\pi} + 2r_ev_e, \\
  \dot v_e &= v_e^2 +\eta_e +J_ew_e\otimes r_e - J_i \tau_i w_i\otimes r_i - \pi^2 r_e^2, \\
  \tau_i^2 \dot r_i &= \frac{\Delta}{\pi} + 2\tau_ir_iv_i, \\
  \tau_i \dot v_i &= v_i^2 +\eta_i +J_ew_e\otimes r_e - J_i\tau_iw_i\otimes r_i - \pi^2 \tau_i^2 r_i^2.
  \end{aligned}
  \label{eq:EImodel}
\end{equation*}
The subscripts $e$, $i$ indicate whether a variable or parameter refers to the
excitatory or inhibitory population, respectively: the two populations have, for
simplicity, the same heterogeneity parameter $\Delta$, but they have possibly different
membrane time constants and average background currents. In single-population mean
fields, excitation and inhibition are artificially lumped into a single
excitatory-inhibitory kernel (see for instance \eqref{eq:expkernel},
\eqref{eq:dampedKernel}, and \eqref{eq:cosKernel}), whereas
in the new, more realistic model the kernels are separate
\begin{equation}
  w_e(x) = \mbox{e}^{-|x|}, \quad w_i(x) = \frac{1}{4} \mbox{e}^{-|x|/2}.
\end{equation}
%%%%%%%%%%%%%%%%%%%%%%%%%%%%%%%%%%%%%%%%%%%%%%%%%%%%%%%%%%%%%%%%%
 \begin{figure*}
\includegraphics[width=0.9\textwidth]{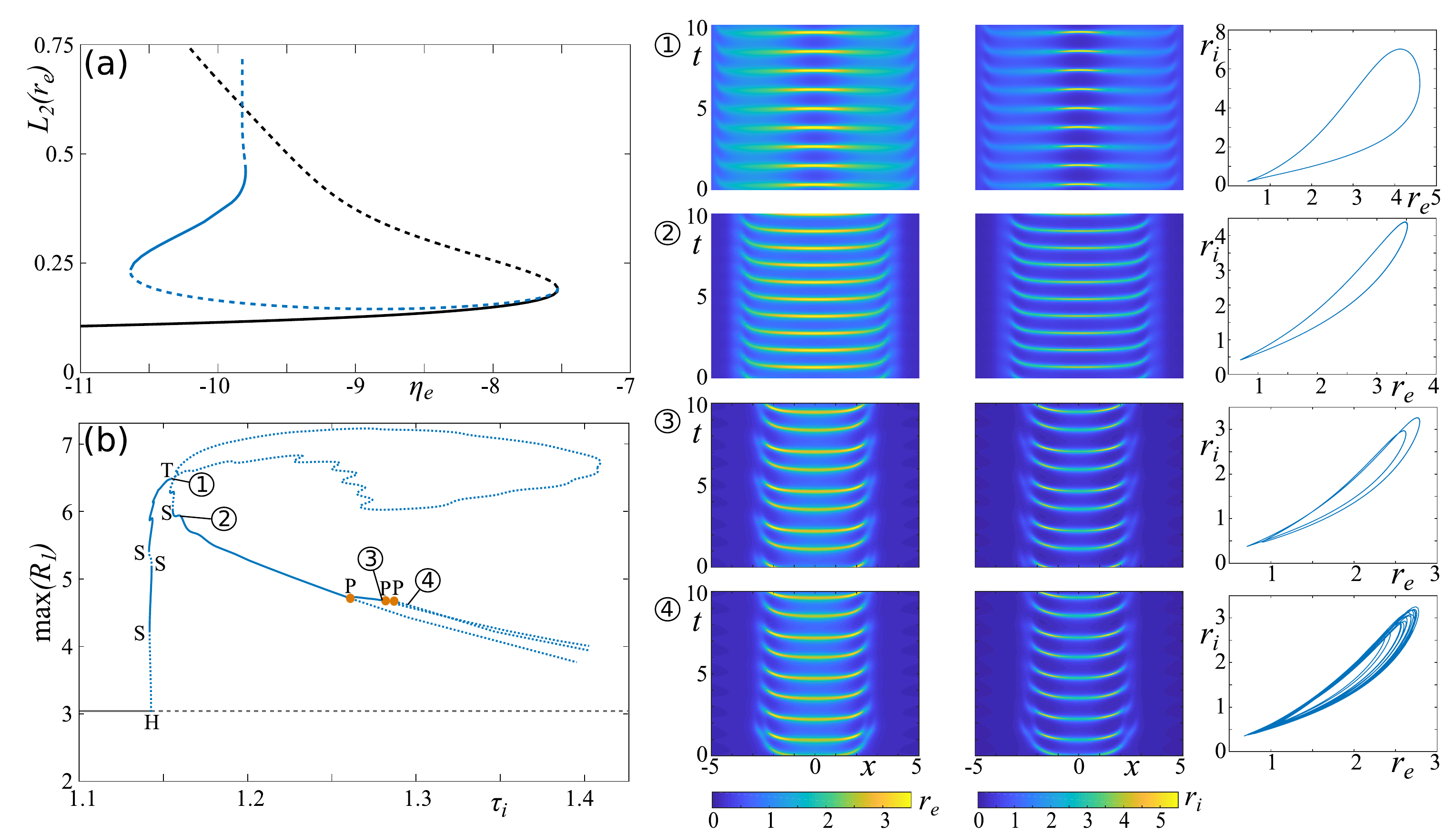}
 \caption{\label{fig16} (a) Bifurcation diagram of bump solutions in the E-I network
   (blue), and spatially uniform solutions (black) for $\tau_i = 1$. (b)
   Bifurcation diagram of emerging limit cycles (showing maxima of $R_1(t)$) using
   $\tau_i$ as bifurcation parameter. H:~Hopf bifurcation, S: saddle node bifurcation, T: torus
   bifurcation, P: period doubling bifurcation.
 Parameters: $\eta_e = \eta_i = -10$, $J_e = J_i = 15\sqrt{\Delta}$, $\Delta = 2$.}
 \end{figure*}
%%%%%%%%%%%%%%%%%%%%%%%%%%%%%%%%%%%%%%%%%%%%%%%%%%%%%%%%%%%%%%%%%
The connectivity parameters are chosen to be $J_e = J_i = J$ to recover a similar
setting used in the lumped model. In Figure~\ref{fig16}a we show the bifurcation
diagram of localized solutions using $\eta_e$ as bifurcation parameter. The
bifurcation structure is similar to the lumped model, with the exception that the
range of parameters for which stable solutions exist is narrower. This computation
confirms that stationary bumps are supported by the two-population network. In order
to hunt for oscillons, we continue the solution for $\eta_e = -10$ in the
parameter $\tau_i$: the bump becomes unstable at a subcritical Hopf bifurcation at
$\tau_i \approx 1.14$, restabilizes at a saddle-node bifurcation, and undergoes a
sequence of saddle-node bifurcations leading to a torus bifurcation (i.e. generalized Hopf bifurcation). 
The branch
eventually restabilizes at a further saddle-node, leading to a period-doubling cascade
which initiates around $\tau_i \approx 1.26$, and to chaos at $\tau_i > 1.29$ (Figure~\ref{fig16}b).

\begin{figure}
\includegraphics[width=0.5\textwidth]{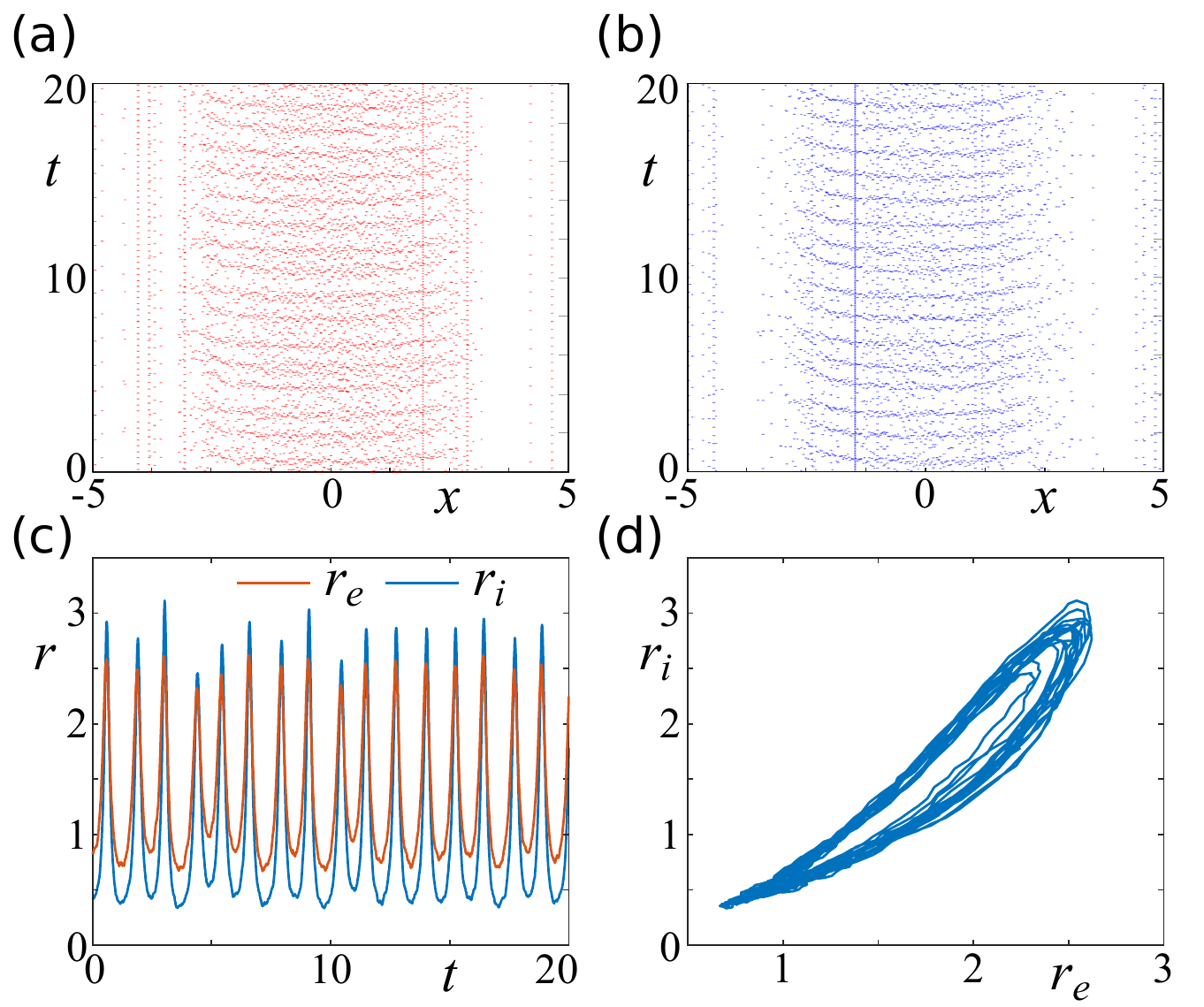}
 \caption{\label{fig20} Chaotic solution in spiking network. 
 (a) Rastergram of excitatory population.
 (b) Rastergram of inhibitory population.
 (c) Excitatory and inhibitory spike rates averaged on interval $-1\!<\!x\!<\!1$ and sliding window in $t$ (width $10^{-3}$).
 (d) Phase portrait of spike rates in (c).
   Parameters: $\eta_e = \eta_i = -10$, $J_e = J_i = 15\sqrt{\Delta}$, $\Delta = 2$, $\tau_i = 1.295$.
   These parameters correspond to point (4) in Figure~\ref{fig16}.}
%   \da[inline]{Refer to the parameters correspond to Figure 10b, point 4 (see comment
%   17).}
%   \HS[inline]{Done.}
 \end{figure}

In Figure~\ref{fig16} we also show numerical examples of a
stable period-doubled solution at $\tau_i = 1.28$ and 
a chaotic solution at $\tau_i = 1.295$. We do not observe
oscillons beyond $\tau_i = 1.3$. Chaotic solutions can also be reproduced in the
spiking network model, see Figure~\ref{fig20}.

%%%%%%%%%%%%%%%%%%%%%%%%%%%%%%%%%%%%%%%%%%%%%%%%%%%%%%%%%%%%%%%%%
%%%%%%%%%%%%%%%%%%%%%%%%%%%%%%%%%%%%%%%%%%%%%%%%%%%%%%%%%%%%%%%%%

%
%
%

\section{Discussion}\label{sec:conclusion}

We introduced a framework to study localized solutions in a neural field model that was
recently derived as an exact representation of the mean field dynamics of networks of
spiking neurons. Although this model does not permit closed-form solutions such as
the Amari model with Heaviside firing rates, we show that it is possible to give an
analytical estimate for the range of model parameters for which stable localized
solutions exist. The structure of the QIF neural field model permits the straightforward use
of Galerkin methods, which unlike the Amari model has a linear nonlocal term.

We have demonstrated that stationary equations can
be transformed into a formulation that is equivalent to the stationary Amari model,
provided an effective firing-rate function is defined. The significance of such a 
firing rate is chiefly mathematical: the neural field possesses a rate variable, which
is combined with the voltage variable in the effective firing rate; however, this
transformation allows to map out patterned steady states of the QIF neural field model using
the same toolkit available for the Amari formulation. In both models localized
solutions emerge subcritically from a branch of homogeneous
steady states, which then restabilize at a saddle node bifurcation. In the Amari
model, this behavior is parametrized by a firing threshold, whereas here we use the
average excitability of the network to map out solutions. However, there is a
correspondence between the excitability of the model used here and the firing
threshold in the Amari model, in the sense that an increase in the firing threshold
in the latter corresponds to a decrease in the excitability in the former. In
addition, techniques developed for piecewise-linear firing rate functions in the
Amari model~\cite{coombes10}, could be adapted to work for steady states in the QIF
neural field model, using the correspondence described above. Furthermore,
all branches of stationary solutions computed in this paper, including
the snaking branches, also occur in standard rate-based models. The crucial
difference lies in the transient dynamics of the two models, which makes the model
considered here dynamically richer and more realistic.

The development of a Galerkin method opened up the possibility to study
oscillons using numerical bifurcation analysis. We focused here on sinusoidal forcing
of bump solutions, which is a proxy of oscillations ubiquitous in neuronal systems.
In previous work~\cite{schmidt18}, the neural mass version of this model was studied
in terms of its response to oscillatory forcing in various frequency bands, and the
present paper makes this exploration feasible also in the spatially-extended model. We
leave this exploration to a future publication.

In coupled networks of excitatory and inhibitory populations,
a small change in the inhibitory membrane time scale can have a significant effect on
the existence and dynamics of bump solutions, and can elicit oscillons. This was
demonstrated for instantaneous
synapses, and it remains to be seen how the dynamics changes when synaptic delays are
introduced to the model.
Interestingly, oscillatory solutions which undergo torus bifurcations have been observed
in spatially extended networks of excitatory and inhibitory neurons with conductance-based
dynamics~\cite{folias10}.
Another natural extension would be to examine coupled
multi-layer neural field models~\cite{bressloff15}, which are known to give rise to
localized bump solutions when neither layer does in isolation~\cite{folias11}.

The Galerkin numerical methods derived in this paper can be applied directly to more
general spatially-extended models of QIF networks, such as the ones mentioned above.
For instance, adding a synaptic variable can be accounted for with an additional
Galerkin expansion, and $n$ scalar variables per additional evolution equation.

Single population, QIF neural mass models with chemical as well as electrical
synapses have recently been developed~\cite{pietras2019}, and it was found that
oscillations originate at Hopf bifurcations. Spatially-extended versions of this
model would then have the possibility of forming oscillons with a single population,
although it is not clear whether Hopf bifurcations of bumps will occur near Hopf
bifurcations of homogeneous states, which are the ones mapped in
Reference~\cite{pietras2019}.

Understanding how slow-fast temporal scales are generated by the discrete network is
an open question, which has recently been addressed in networks of sparsely-coupled
networks of QIF neurons \cite{Bi2019}. Employing our numerical methodology to 
these
macroscopic mean fields is also possible, and one could study how such
slow-fast phenomena occur in more realistic, spatially-extended networks.

\subsection*{Acknowledgments}

HS acknowledges financial support from the Spanish Ministry of Economics and
Competitiveness through the Mar\'ia de Maeztu Programme for Units of Excellence in
R\&D (MDM-2014-0445) and grant MTM2015-71509-C2-1-R, and from the German Research
Council (DFG (KN 588/7-1) within priority programme `Computational Connectomics' (SPP
2041) ).

\end{document}